\documentclass[]{interact}
\usepackage{amsmath}
\usepackage{amsfonts}
\usepackage{amssymb}
\usepackage{revsymb}
\usepackage{mathtools}
\usepackage{graphicx}
\usepackage{mhchem}
\usepackage{siunitx}
\usepackage{bm}
\usepackage{dcolumn}
\usepackage{multirow}

\usepackage{tikz}
\usepackage{pgfplots}
\usepackage{makecell}
\usetikzlibrary{shapes.arrows}
\usetikzlibrary{shapes.geometric}
\usetikzlibrary{positioning}
\usetikzlibrary{calc}
\usetikzlibrary{decorations.pathmorphing}
\usetikzlibrary{fit}
\pgfdeclarelayer{background}
\pgfdeclarelayer{foreground}
\pgfsetlayers{background,main,foreground}
\usetikzlibrary{automata}

\usepackage[numbers,sort&compress,merge]{natbib}
\bibpunct[, ]{(}{)}{,}{n}{,}{,}
\renewcommand\bibfont{\fontsize{10}{12}\selectfont}
\renewcommand\citenumfont[1]{\textit{#1}}
\renewcommand\bibnumfmt[1]{(#1)}

\newcommand{\sref}[1]{sec.~\ref{#1}}
\newcommand{\fref}[1]{fig.~\ref{#1}}
\newcommand{\eref}[1]{eq.~(\ref{#1})}

\MHInternalSyntaxOn
\newcommand{\adjintertext}[3]
{\ifvmode\else\\\@empty\fi
  \noalign{%
    \vskip-\lineskiplimit      
    \vskip\normallineskiplimit 
    \vskip#1
     \vbox{\normalbaselines
       \ifdim
         \ifdim\@totalleftmargin=\z@
           \linewidth
         \else
           -\maxdimen
         \fi
       =\columnwidth
      \else \parshape\@ne \@totalleftmargin \linewidth
      \fi
      \noindent#3\par}%
    \vskip-\lineskiplimit      
    \vskip\normallineskiplimit 
    \vskip#2
 }}%
\MHInternalSyntaxOff

\def\ket#1{|\,#1 \,\rangle}
\def\bra#1{\langle \, #1 \,|}

\def\vec#1{\mathbf{#1}}

\begin{document}
\title{Laser spectroscopy of $\mathbf{^{176}Lu^+}$}
\author{\name{R.~Kaewuam, A. Roy, T. R. Tan, K. J. Arnold, and M. D. Barrett}
\affil{Center for Quantum Technologies, 3 Science Drive 2, Singapore, 117543;\\Department of Physics, National University of Singapore, 2 Science Drive 3, Singapore, 117551}}
\maketitle
\begin{abstract}
We perform high resolution spectroscopy on $^{176}$Lu$^+$ including the $^1S_0\leftrightarrow{^3}D_1$ and  $^1S_0\leftrightarrow{^3}D_2$ clock transitions.  Hyperfine structures and optical frequencies relative to the $^1S_0$ ground state of four low lying excited states are given to a few tens of kHz resolution.  This covers the most relevant transitions involved in clock operation with this isotope.  Additionally, measurements of the $^3D_2$ hyperfine structure may provide access to higher order nuclear moments, specifically the magnetic octupole and electric hexadecapole moments.
\end{abstract}
\section{Introduction}
Singly-ionized lutetium has recently been proposed as a promising clock candidate with the $^3D_1$ and $^3D_2$ excited states providing suitably long-lived clock states.  As discussed in \cite{MDB1}, averaging over hyperfine levels eliminates shifts associated with the electronic angular momentum, $J$, realizing an effective $J=0$ level.  The even isotope $^{176}$Lu$^+$ has $m=0$ states which, together with large hyperfine and fine structure splittings, provides insensitivity to magnetic fields for the clock transitions and the hyperfine-averaged reference.  The $^3D_1\leftrightarrow{^3}P_0$ transition has a linewidth of $\sim2.5\,\mathrm{MHz}$, which provides a low Doppler cooling limit while maintaining sufficient scattering for state detection.  Theoretical calculations of the differential DC scalar polarizabilities indicate the possibility of a negative value with a low blackbody radiation shift.  A negative value would further allow clock shifts from excess micromotion to be cancelled by a judicious choice of trap drive frequency \cite{micromotion, MDB2}.  This has lead to the recent consideration of clock operation with many ions \cite{MDB2, MDB3}.

Basic atomic properties have been experimentally explored \cite{LuProperties}, and the first observation of the $^1S_0\leftrightarrow{^3}D_1$ clock transition has also been reported \cite{Lu175Clock}.  These initial experiments were carried out with $^{175}$Lu$^+$ owing to its more prominent abundance.  This made initial spectroscopy more readily accessible with some results already available in the literature, albeit with modest accuracy \cite{ref1,ref2,ref3,ref4}.  In contrast, very little was known about the preferred $^{176}$Lu$^+$ isotope.  Here we provide a complete energy spectrum for all levels of $^3D_1$, $^3D_2$, $^3P_0$, and $^3P_1$ relative to the $^1S_0$ ground-state.  This provides the most important frequencies for clock operation and improves on the initial work reported in \cite{Samuel}.  
\section{Experimental Setup}
\label{sectOne}
Experiments are performed in a linear Paul trap similar in design to those used in previous work \cite{LuProperties, Lu175Clock}. It consists of two axial endcaps separated by $2\,\mathrm{mm}$ and four rods arranged on a square with sides $1.2\,\mathrm{mm}$ in length. All electrodes are made from $0.45\,\mathrm{mm}$ electropolished copper-beryllium rods.  Radial confinement is provided by a $16.8\,\mathrm{MHz}$ radio-frequency (rf) potential applied to a pair of diagonally opposing electrodes via a helical quarter-wave resonator.  A dc voltage applied to the other pair of diagonally opposing electrodes ensures a splitting of the transverse trapping frequencies.  The endcaps are held at $8\,\mathrm{V}$ to provide axial confinement.  The measured trap frequencies of a single $\mathrm{Lu}^+$ are $(\omega_x,\omega_y,\omega_z) \approx 2\pi\times(608,560,134)$ kHz, with the trap axis along $z$.  A quantization axis is defined by a $\sim0.2\,\mathrm{mT}$ magnetic field oriented perpendicular to the trap axis. 

\begin{figure}
\begin{tikzpicture}[scale=1.0,
      level/.style={line width=2pt,line cap=round},
      trans/.style={line width=1pt,color=black,<->,shorten >=3pt,shorten <=3pt,>=stealth},
      label/.style = {midway,align=center,color=black,fill=white},
      dimension/.style ={line width=0.75pt,color=black,|-|},
      laser/.style ={line width=2pt,color=black,<-}
    ]
    
    \coordinate (p1) at (2.0cm,5cm);
    \coordinate (d1) at (4.0cm,2.3cm);    
    \coordinate (s1) at (0.2cm,0cm);  
    \coordinate (lev) at (1.5cm,0cm);    

\draw[level] (s1) node[left] {}  -- ++(lev) node[midway] (seps1) {}  node[left,xshift=-1.5cm] {$^1$S$_{0}$};
\draw[level] (p1) node[left] {}  -- ++(lev) node[midway] (sepp0) {} node [left,xshift=-1.5cm] {$^3$P$_{0}$};
\draw[level] ($(p1)+(0,1.2cm)$) node[left] {}  -- ++(lev) node[midway] (sepp1) {} node [left,xshift=-1.5cm] {$^3$P$_{1}$};
\draw[level] (d1)   -- ++(lev) node[midway] (sepd1) {} node[right] {$^3$D$_{1}$};
\draw[level] ($(d1)+(0,0.8cm)$) -- ++(lev) node[midway] (sepd2) {}  node[right] {$^3$D$_{2}$};
\normalsize
\draw[trans] ([xshift=-0.4cm]seps1.center) --([xshift=-0.2cm]sepp1.center) node[label,midway]{ 350\,nm};
\draw[trans] ($(sepd2)+(0.5cm,0)$) --($(sepp1)+(0.5cm,0)$) node[label,xshift=0.2cm,yshift=0.2cm]{ 622\,nm};
\draw[trans] ([xshift=0.3cm]sepp0.center) --([xshift=0.1cm]sepd1.center) node[label,midway,yshift=0.0cm,xshift=-0.0cm]{646\,nm};
\draw[trans] ([xshift=0.3 cm]seps1.center)  --($(sepd1)+(0cm,0)$) node[label,xshift=0.4cm]{ 848\,nm};
\draw[trans] ([xshift=-0.1 cm]seps1.center)  --($(sepd2)+(-0.5cm,0)$) node[label,yshift=0.35cm,xshift=0.2cm]{ 804\,nm};

\coordinate (tc) at (10.5cm,2.5cm);

\filldraw[gray] ($(tc)+(1.65cm,0)$) -- ++ (0.175cm,0.175cm) -- ($(tc)+(2.5cm,0.175)$) -- ($(tc)+(2.5cm,-0.175)$) --($(tc)+(1.825cm,-0.175)$) -- ($(tc)+(1.65cm,0)$) ;
\filldraw[gray] ($(tc)+(-1.65cm,0)$) -- ++ (-0.175cm,0.175cm) -- ($(tc)+(-2.5cm,0.175)$) -- ($(tc)+(-2.5cm,-0.175)$) --($(tc)+(-1.825cm,-0.175)$) -- ($(tc)+(-1.65cm,0)$) ;
\filldraw[gray] ($(tc)+(-2.5cm,0.65cm)$) -- ++(0cm,0.35cm) --++(5cm,0cm)-- ++(0cm,-0.35cm) -- ($(tc)+(-2.5cm,0.65cm)$);
\filldraw[gray] ($(tc)+(-2.5cm,-0.65cm)$) -- ++(0cm,-0.35cm) --++(5cm,0cm)-- ++(0cm,0.35cm) -- ($(tc)+(-2.5cm,-0.65cm)$);    

\filldraw[black] (tc) circle (0.075cm) node[left] {\footnotesize$^{176}$Lu$^+$ };


\draw[laser] ($(tc)+(0cm,1.2cm)$) -- ++ (0cm,1.2cm) node[right,yshift=0.4cm,xshift=-1.6cm] {$646, 804, 848\;\mathrm{nm}~(\pi)$};
\draw[laser] ($(tc)+(1.2cm,1.2cm)$) -- ++ (0.7cm,0.7cm) node[right,yshift=0.2cm,xshift=-1.1cm] {$646\;\mathrm{nm}~(\perp)$};
\draw[laser] ($(tc)+(-1.2cm,1.2cm)$) -- ++ (-0.7cm,0.7cm) node[right,yshift=0.2cm,xshift=-1.1cm] {$350\;\mathrm{nm}~(\perp)$};
\draw[laser] ($(tc)+(1.2cm,-1.2cm)$) -- ++ (0.7cm,-0.7cm)  node[right,yshift=-0.3cm,xshift=-1.1cm] {$622\;\mathrm{nm}~(\perp)$};

\node (bf) at ($(tc)+(-1.2cm,-1.75cm)$) {\Large \raisebox{1pt}{$\otimes$} $\vec{B}$};

\node (a) at (-0.5cm,6.2cm) {\large (a)};
\node (b) at ($(tc)+(-3.3cm,3.7cm)$) {\large (b)};

\end{tikzpicture}
 \caption{(a) Low-lying level structure of Lu$^+$ showing the transitions relevant to this work, (b) Schematic of trap setup showing the orientations and polarizations of the laser fields and the magnetic field direction.  Polarization parallel (orthogonal) to the magnetic field direction is indicated by $\pi\,(\perp)$.}
 \label{figure1}
\end{figure}
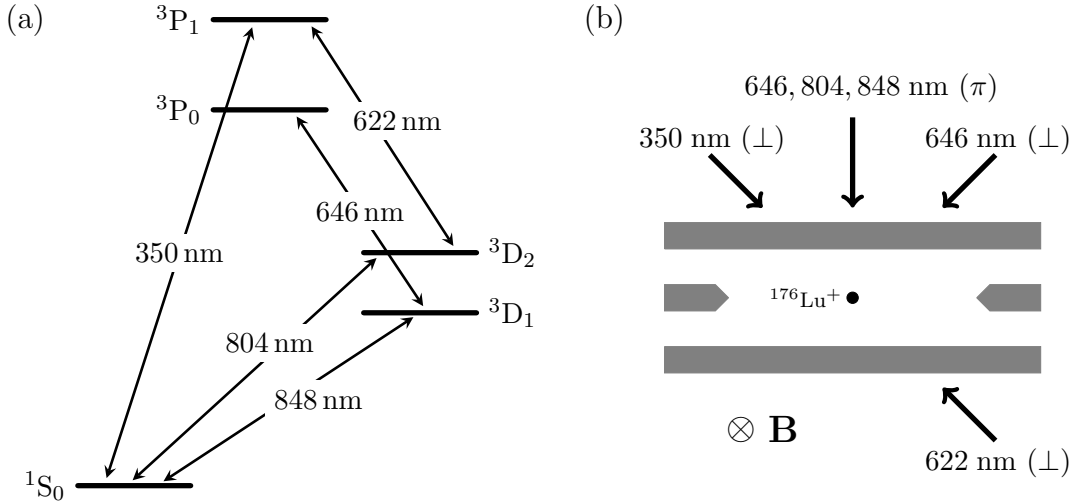

The low-lying level structure of $^{176}\mathrm{Lu}^+$ is shown in \fref{figure1}(a) and shows the transitions relevant to this work.  There are two clock transitions: a highly forbidden magnetic dipole (M1) transition $^1S_0\leftrightarrow{}^3D_1$ at $848\,\mathrm{nm}$ with an estimated lifetime of 172 hours \cite{Lu175Clock}, and a spin-forbidden electric quadrupole (E2) transition $^1S_0\leftrightarrow{}^3D_2$ at $804\,\mathrm{nm}$ with a measured lifetime of $17.3\,\mathrm{s}$ \cite{LuProperties}.  Detection, cooling, and state preparation are achieved via scattering on the $^3D_1\leftrightarrow{}^3P_0$ transition at $646\,\mathrm{nm}$.  Optical pumping to $^3D_1$ is achieved via lasers addressing the $^1S_0\leftrightarrow{}^3P_1$ and $^3D_2\leftrightarrow{}^3P_1$ transitions at  350 and 622 nm, respectively.

The primary $848$-nm clock laser is an extended-cavity diode laser (ECDL) with an interference-filter feedback similar in design to that reported in \cite{LaserSetup}. The laser is stabilized to a $3.5\,\mathrm{kHz}$ linewidth high finesse ($\mathcal{F} \approx 400000$) reference cavity and its frequency is tuned near to the $^1S_0(F=7)\leftrightarrow{}^3D_1(F=7)$ transition. To obtain the required optical power for interrogating the narrow transitions, an additional laser diode is injection-locked to the primary laser.  The laser light is delivered to the experiment through a single-mode fibre.  As shown in \fref{figure1}(b), it propagates orthogonal to the magnetic field direction with $\pi$-polarization. This configuration allows $\Delta m_F=\pm 1$ M1 transitions to be driven. 

The 804-nm clock laser is a grating-stabilized ECDL with a $9\,\mathrm{cm}$ long external cavity. The laser is locked to a $10\,\mathrm{cm}$ long optical cavity with a linewidth of ${\sim100}\,\mathrm{kHz}$. This laser is coupled to the same fiber as the 848-nm clock lasers with the same polarization.  This also allows $\Delta m_F=\pm 1$ E2 transitions to be driven.  The laser frequency is controlled by a wide-band electro-optic modulator (EOM) which offsets the lock to the cavity.

The 646-nm laser is a grating-stabilized ECDL locked to an optical transfer cavity, which is referenced to the 848-nm clock laser. To couple the three separate hyperfine levels of $^3D_1$ to the upper $^3P_0$ states, a wideband EOM generates sidebands of $\sim10.9\,\mathrm{GHz}$.  The sidebands address the $F=6$ and $8$ hyperfine levels; and a separate carrier beam is shifted by a double-passed acousto-optic modulator (AOM) to address the remaining $F=7$ level. The beams are then combined into a single-mode fiber.  This light is linearly-polarized orthogonal to the  magnetic field direction ($\perp$) to avoid any dark states during detection and cooling.  An additional $\pi$-polarized carrier beam is coupled into the same fiber as the clock laser beams.  The $\pi$-polarized beam, together with the orthogonally-polarized sidebands, provides optical pumping into $\ket{{}^3D_1,F=7,m_F=0}$.  Sideband frequencies and carrier offsets were chosen based on previous lower resolution measurements \cite{Samuel}.

The 350-nm laser is a frequency-doubled grating-stabilized ECDL. The fundamental at $701\,\mathrm{nm}$ is locked to an optical transfer cavity, which is referenced to a cesium vapor cell via a diode laser at $852\,\mathrm{nm}$. A double-pass AOM is used to provide a tunable offset of the 701-nm laser from the transfer cavity.  The fundamental is frequency doubled with a BBO crystal inside a custom-built bow-tie cavity.  At the ion, the 350-nm light is $\perp$-polarized as illustrated in \fref{figure1}(b).

The 622-nm laser is also a frequency-doubled grating-stabilized ECDL.  The fundamental is frequency doubled using a fiberized, periodically-poled KTP waveguide.  The 622-nm light is coupled to a wide-band EOM which generates multiple sidebands to couple all five levels of $^3D_2$ to $^3P_1$.  As with the 646-nm laser, sideband frequencies were based on previous lower resolution measurements \cite{Samuel}. The 622-nm carrier frequency is stabilized using a wavemeter with a specified $\sim$10 MHz accuracy.  At the ion, the 622-nm light is also $\perp$-polarized as illustrated in \fref{figure1}(b).
\section{Measurements}
\label{sectTwo}
Optical frequency measurements carried out in this work are obtained by referencing the lasers to an optical frequency comb.  The frequency comb is stabilized to a GPS-disciplined rf oscillator, which has a specified $\lesssim 5\times 10^{-13}$ accuracy.  Frequency measurements of the clock laser on different days are consistent with the specification.  We therefore use this value in determining the uncertainties in optical frequency measurements.
\subsection{$^3D_1$}
\label{3D1}
Frequency references for  $^1S_0\leftrightarrow{^3}D_1$ transitions are denoted $f_{1,F}$, where $F$ denotes the hyperfine level of $^3D_1$.  These frequencies are determined by a combination of optical and microwave spectroscopy.  Optical spectroscopy starts with driving clock transitions from $\ket{^3D_1,7,0}$ to $\ket{^1S_0,7,\pm1}$.  After Doppler cooling, the ion is optically pumped to $\ket{^3D_1,7,0}$.    A $12.5\,\mathrm{ms}$ interrogation pulse then drives the atom down to either $\ket{^1S_0,7,\pm1}$ which is followed by detection.  Using the technique discussed in \cite{SrIonLock}, the clock laser is servoed to both $^1S_0$ Zeeman states to stabilize the laser frequency to the center of the ground-state Zeeman spectrum.  Referencing the laser to the frequency comb gives 
\begin{equation}
f_{1,7}=353\,638\,527\,912\,390\,(200)\,\mathrm{Hz},
\end{equation}
for a $\sim 7000\,\mathrm{s}$ integration time. The uncertainty is limited by the comb as systematic shifts arising from the quadratic Zeeman, quadrupole, and probe-induced AC stark shifts are estimated to give a combined shift of less than $10\,\mathrm{Hz}$.

\begin{figure}[hbt]
\begin{center}
  \includegraphics{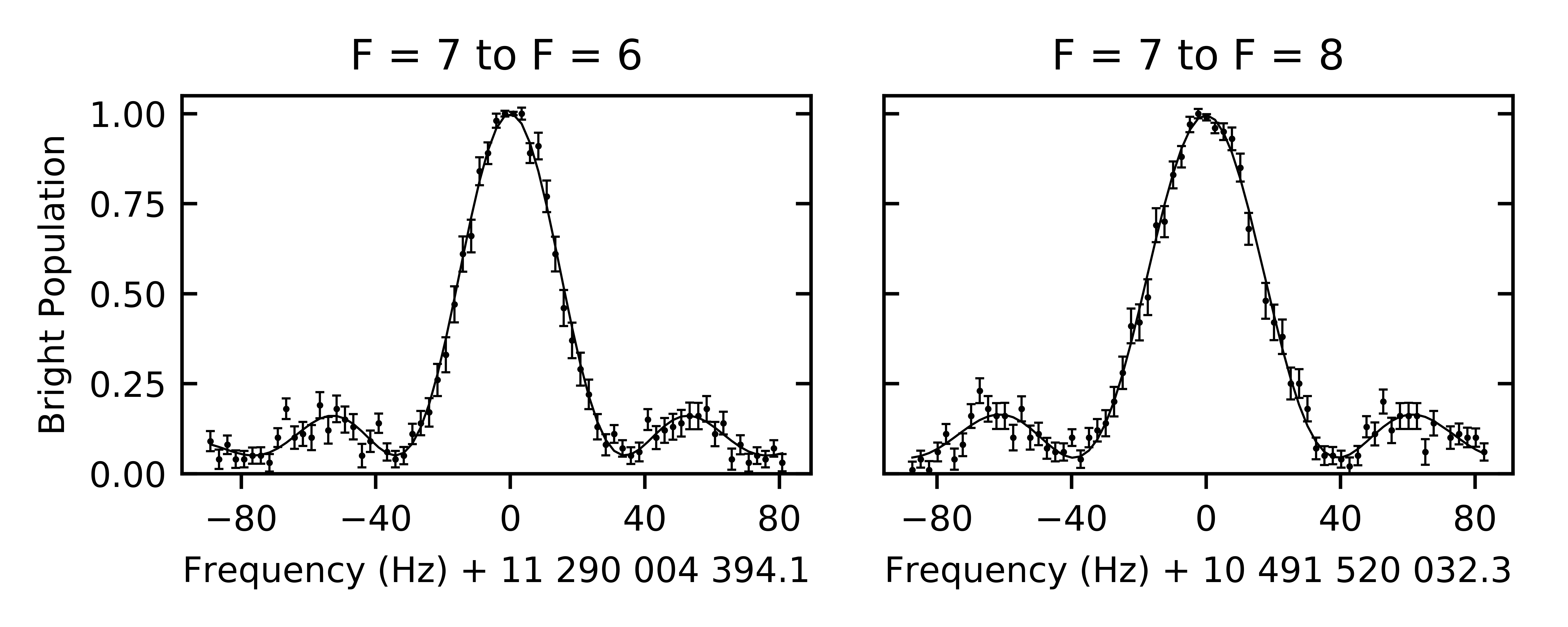}
  \caption{Microwave transitions between $m_F=0$ hyperfine states of $^3D_1$: $F=7\leftrightarrow F=6$ (left) and $F=7\leftrightarrow F=8$ (right).  Frequencies are given relative to the measured line centers.}
  \label{MicrowaveScan}
\end{center}
\end{figure}

The clock transition provides shelving from $\ket{^3D_1,7,0}$ to $\ket{^1S_0,7,-1}$, which allows state-sensitive detection and microwave spectroscopy of the $^3D_1$ manifold.  Following Doppler cooling, the ion is optically pumped to $\ket{^3D_1,7,0}$.  A microwave pulse then transfers population to either $\ket{^3D_1,6,0}$ or $\ket{^3D_1,8,0}$.  Any remaining population in $\ket{^3D_1,7,0}$ is then shelved to  $\ket{^1S_0,7,-1}$ prior to detection.  The probability the ion remains in $^3D_1$ is determined from 200 cycles of the experiment and resulting scans of the microwave frequency for each transition are given in \fref{MicrowaveScan}. 

Center frequencies extracted from each of the microwave scans have a statistical uncertainty $\sim1\,\mathrm{Hz}$.  At this level, quadratic Zeeman shifts of the $m=0$ states must be accounted for.  The magnetic field can be determined by either the Zeeman splitting of the ground states inferred from the servo data of the clock transition, or from the difference frequency of microwave fields driving $\Delta m_F=\pm1$ transitions in $^3D_1$.  Using the ground-state $g$-factor $g_I=-2.4360(34)\times 10^{-4}$ as measured for neutral lutetium \cite{LuIgI} and $g_J=0.5$ for $^3D_1$, the magnetic fields derived by either approach agree to better than $1\%$.  We conservatively take the field to be $0.207(2)\,\mathrm{mT}$, as inferred from the ground-state Zeeman splitting.  The Zeeman-corrected hyperfine splittings are then found to be
\begin{subequations}
\begin{align}
f_{1,6}-f_{1,7}&=11\,290\,004\,289.0(2.6)\,\mathrm{Hz}\\
\adjintertext{0pt}{0pt}{and}
f_{1,7}-f_{1,8}&=10\,491\,519\,944.7(2.5)\,\mathrm{Hz}.
\end{align}
\end{subequations}
Details of the Zeeman corrections are given in appendix~\ref{3D1Zeeman}.  The given hyperfine splittings have not been corrected for quadrupole shifts, which we estimate to be $\lesssim 1\,\mathrm{Hz}$ based on the measured trap frequencies, or for AC Zeeman shifts, which arise from off-resonant coupling to $m=\pm1$ states by the circular polarization components of the microwave field.  Measuring the Rabi frequencies for $\Delta m_F=\pm1$ transitions allows us to estimate the polarization components of the microwave fields from which we infer shifts well below $1\,\mathrm{Hz}$.

Driving the $^1S_0\leftrightarrow{}^3D_1$ clock transition provides the means to prepare the ion in $\ket{^1S_0,7,\pm1}$ with probability $p_s\sim 95\%$, which is limited by both state preparation of $\ket{^3D_1,7,0}$ and the clock $\pi$-pulse.  This can be conditionally improved by state detection after transfer to the ground state. If the ion is detected bright, state preparation of $\ket{^3D_1,7,0}$ and shelving to $\ket{^1S_0,7,\pm1}$ is repeated.  When a dark state is confirmed, the experiment proceeds.  This conditional state preparation improves the fidelity of $\ket{^1S_0,7,\pm1}$ to better than $99.9\%$ and is used as a starting point for measurements of other levels.
\subsection{$^3D_2$}
Frequency references for  $^1S_0\leftrightarrow{^3}D_2$ transitions are denoted $f_{2,F}$, where $F$ denotes the hyperfine level of $^3D_2$.  These frequencies are determined by optical spectroscopy.  The ion is first conditionally prepared in $\ket{^1S_0,7,\pm1}$ and then driven to $\ket{^3D_2,F',0}$ using the 804-nm clock laser.  For each upper state, the clock-laser intensity is adjusted to give a $\sim 500\,\mathrm{\mu s}$ resonant $\pi$ time, which is limited by the laser coherence time.  This provides sufficient resolution to servo the 804-nm laser frequency to the center of the ground state Zeeman spectrum as for the 848-nm case.  The 804-nm servo loop is interleaved with that stabilizing the 848-nm frequency. Comparisons with the frequency comb then give the optical frequencies for each transition.

The main systematic shifts arise from the quadratic Zeeman shift of the upper levels.  To determine the magnetic field, we measure the ground-state Zeeman splitting between the $m_F=\pm1$ states and estimate the field as for the $^3D_1$ case.  Since the polarization of the 804-nm clock laser couples both $m_F=\pm1$ states to the upper $m_F=0$ level, there is a shift in one transition due to off-resonant coupling to the other.  Being equal and opposite for the two transitions, this cancels for the Zeeman-averaged frequency but degrades the accurate determination of the magnetic field.  We therefore determine the magnetic field from the 848-nm servo where the coupling is much smaller and the effect is negligible.  Using the inferred magnetic field of $0.207(2)\,\mathrm{mT}$, the Zeeman-corrected optical frequencies $f_{2,F}$ are
\begin{subequations}
\label{f2F}
\begin{align}
f_{2,5}&=372\,776\,905\,829\,552\,(200)\,\mathrm{Hz}\label{f25},\\
f_{2,6}&=372\,784\,362\,667\,641\,(200)\,\mathrm{Hz}\label{f26},\\
f_{2,7}&=372\,793\,515\,721\,790\,(200)\,\mathrm{Hz}\label{f27},\\
f_{2,8}&=372\,804\,577\,481\,195\,(200)\,\mathrm{Hz}\label{f28},\\
\adjintertext{0pt}{0pt}{and}
f_{2,9}&=372\,817\,792\,702\,607\,(200)\,\mathrm{Hz}\label{f29},
\end{align}
\end{subequations}
where the uncertainties are limited by the frequency comb. Details of the Zeeman corrections are given in appendix~(\ref{3D2Zeeman}). Quadrupole shifts from the DC confinement fields and AC stark shifts from the probe are estimated to be $\lesssim 1\,\mathrm{Hz}$.

It has been noted that measurement of the hyperfine structure (HFS) for a $J=2$ fine-structure level gives access to higher nuclear moments, specifically the magnetic octupole and the electric hexadecapole moments \cite{Beloy}.  Hyperfine intervals are conventionally parameterized in terms of the HFS constants $A, B,\dots$ each being proportional to relevant nuclear multipole moments.  From relativistic theory of the hyperfine interaction, the HFS constants can be expressed as linear combinations of the measured hyperfine splittings together with second-order correction terms \cite{Beloy}.  Dominant corrections are due to dipole-dipole, and dipole-quadrupole interactions but these terms do not contribute to the HFS $D$ constant, which is proportional to the nuclear electric hexadecapole moment.  This can be determined directly from the measured optical frequencies, $f_{2,F}$, and we find
\begin{align}
\label{Dcoeff}
D&=\frac{11}{214200}\left(204 f_{2,5}-663 f_{2,6}+810 f_{2,7}-442 f_{2,8}+91 f_{2,9}\right)\\
&= -1.824\,(12) \,\mathrm{kHz}.\nonumber
\end{align}
Here the error is completely dominated by the uncertainty in $f_{2,F}$ but this can be reduced with improvements to the comb or direct measurements of the hyperfine splittings via microwave spectroscopy.  We note that the expression for $D$ has only taken into account leading order dipole-dipole, and dipole-quadrupole interactions.  More generally, Woodgate derived the expression \cite[Eq. B5]{Woodgate}
\begin{align}
\frac{\delta A_K(J)}{\begin{pmatrix}I & K & I\\-I & 0 & I\end{pmatrix}\begin{pmatrix}J & K & J\\-J & 0 & J\end{pmatrix}}&=\sum_{k_1,k_2}(-1)^{2(I+J)+k_1+k_2+K}(2K+1)\nonumber\\
&\quad \times \begin{Bmatrix}k_1 & k_2 & K\\I & I & I\end{Bmatrix}\bra{I}|T_n^{k_1}|\ket{I}\bra{I}|T_n^{k_2}|\ket{I}\nonumber\\
&\quad \times \sum_{J'}\begin{Bmatrix}k_1 & k_2 & K\\J & J & J'\end{Bmatrix}\frac{\bra{J}|T_e^{k_1}|\ket{J'}\bra{J'}|T_e^{k_2}|\ket{I}}{E_J-E_{J'}},
\end{align}
which includes all correction terms to second order.  In this equation $\delta A_K$ is related to the appropriate nuclear moment. For $K=4$, $\delta A_4(J)=\delta D$, and this correction should be added to the left hand side of Eq~(\ref{Dcoeff}).  Proper assessment of the hexadecapole moment would thus require calculations of quadrupole-quadrupole ($k_1=2=k_2$) and dipole-octupole ($k_1=1, k_2=3$) corrections.  These were omitted in the treatment given in \cite{Beloy} but may well be significant compared to the value given in Eq~\ref{Dcoeff}.  
\subsection{$^3P_0$}
\label{3P0}
A frequency reference, $f_0$, for the $^3D_1\,(F=7)\leftrightarrow{^3}P_0\,(F=7)$ transition is determined by measuring the rate at which $\ket{^3D_1,7,0}$ is depopulated by a single 646-nm laser beam as a function of the laser frequency.  Determining a depumping rate in such a multi-level system can be complicated, due to multiple scattering back into $\ket{^3D_1,7,0}$ and unwanted population of other states.  Both effects can lead to deviations from a simple exponential decay of the $\ket{^3D_1,7,0}$ population and deviations from a Lorentzian line-profile.    

To eliminate population of other $^3D_1$ states, the ion is conditionally prepared in $\ket{^1S_0,7,-1}$ and then transferred back to $\ket{^3D_1,7,0}$ with a $\pi$-pulse from the 848-nm laser.  This prepares the ion in $\ket{^3D_1,7,0}$ with probability $p_s$ and ensures no other $^3D_1$ states are occupied.  Depumping is achieved using light that is linearly polarized perpendicular to the magnetic field direction.  This symmetrically excites the atom to $\ket{^3P_0,7,\pm1}$ and eliminates any shift of the line-center arising from the Zeeman broadening of the $^3P_0$ level.  This configuration also leads to a simple exponential decay of the $\ket{^3D_1,7,0}$ population, which was confirmed both experimentally and by simulation of the multi-level system.  After a depumping time, $\tau$, $\ket{^3D_1,7,0}$ is shelved to $\ket{^1S_0,7,1}$ which prevents deshelving of population in $\ket{^1S_0,7,-1}$ from state-preparation.  The probability, $p$, that subsequent detection yields a bright state is given by
\begin{equation}
\label{3P0pumping}
p=p_s\left(1-p_s e^{-R \tau}\right),
\end{equation}
where $R$ is the depumping rate.

Experimentally, $p$ is determined from 200 cycles of the experiment, and this is repeated 50 times to determine the statistical variation.  For each detuning, we choose $\tau$ so that the measured value of $p$ is near 0.5 and we infer $R$ by inverting \eref{3P0pumping}.  The value of $p_s$ is determined from independent shelving experiments with sufficient averaging such that the uncertainty in $R$ is dominated by the uncertainty in $p$, which is consistent with expected projection noise.  

\begin{figure}[hbt]
\begin{center}
  \includegraphics{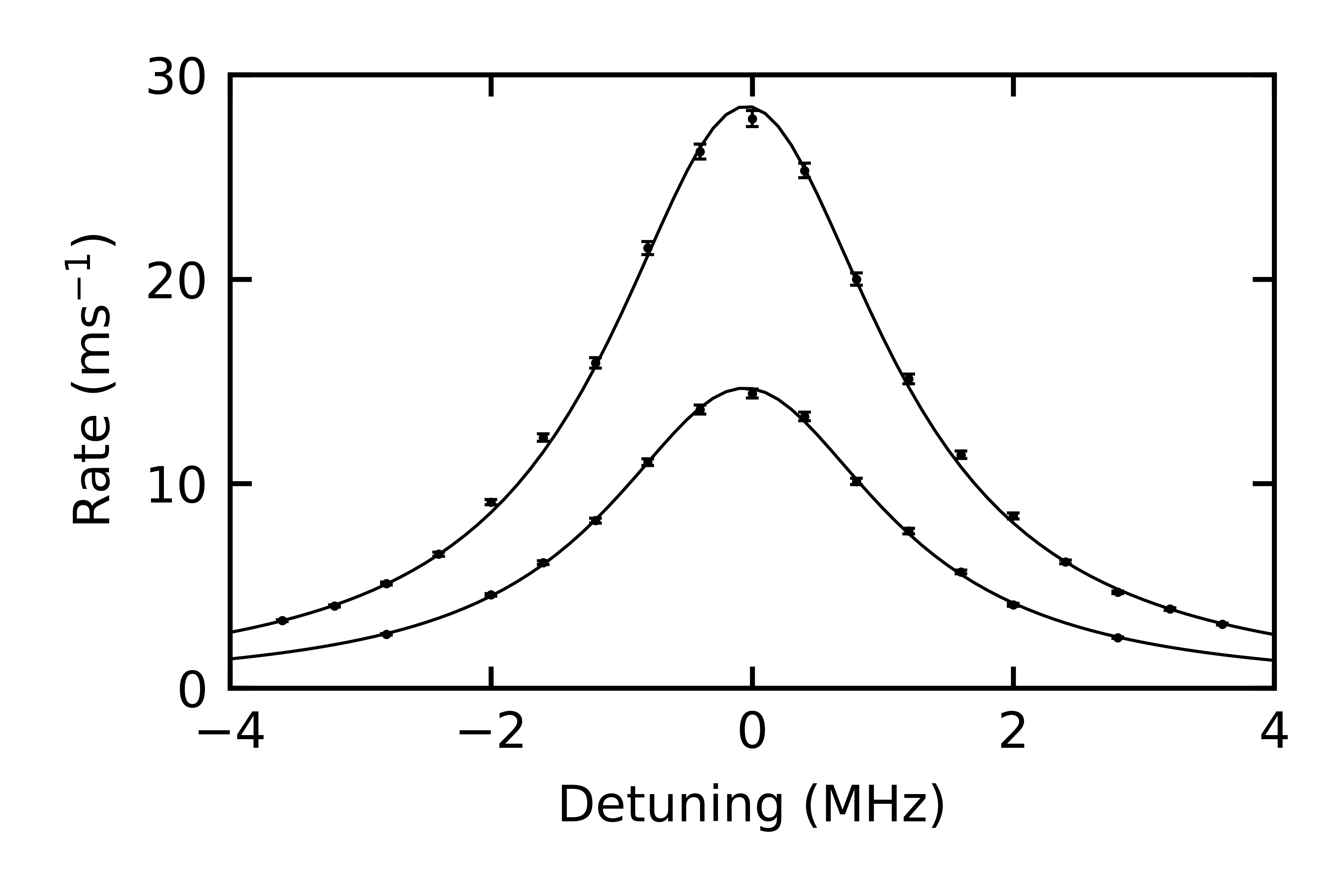}
  \caption{Experimental line shape for the $^3D_1(F=7)\leftrightarrow{^3}P_0(F=7)$ transition for two different pumping intensities. The solid lines are corresponding fits to a Lorentzian profile.}
  \label{3P0Data}
\end{center}
\end{figure}

In \fref{3P0Data}, the inferred values of $R$ as a function of detuning for two different powers of the depumping beam are given.   In both cases the detuning is relative to a conveniently chosen set point of the frequency driving the AOM, which offsets the optical frequency from the lock point. The solid curves are chi-square fits to a Lorentzian.  From fits to the upper data set, the line center relative to the set point and the linewidth are given by $-43.7(7.4)\,\mathrm{kHz}$ and $2\pi\times2.573(17)\,\mathrm{MHz}$, respectively. Corresponding results for the lower data set are $-56.3(7.0)\,\mathrm{kHz}$ and $2\pi\times2.586(21)\,\mathrm{MHz}$.  Errors given are the statistical error derived from the fits.

From measurements against the comb, the 646-nm laser frequency at the lock point of the transfer cavity drifts $\sim100\,\mathrm{kHz}$ over timescales of a few hours, which is comparable to the timescale of the data collection.  As it was not possible at the time to monitor the optical frequency during data collection, reference measurements were taken before and after the collection of each data set.  For each data set, we take the mean of the two reference measurements for the optical frequency at the lock point with the error given by half the difference.  This gives optical frequencies at the line centers for each data set that agree to within $7\,\mathrm{kHz}$.  Taking the average value of the two centers as an estimate for $f_0$ gives
\begin{equation}
f_0=463\,723\,160\,779(30)\,\mathrm{kHz},
\end{equation}
where the error is primarily due to drift of the laser frequency.

There is also good agreement in the fitted linewidths demonstrating that power broadening is not significant as is consistent with the measured depumping rates.  However, laser linewidth and Doppler effects cannot be neglected.  By virtue of the fact that the laser is locked to a cavity, we expect the laser linewidth to be below $1\%$ of the $3\,\mathrm{MHz}$ linewidth of the cavity.  Thus broadening due to the laser is, at most, on the order of a few times the statistical uncertainty of the fit.  In the limit of unresolved motional sidebands, the line profile in the presence of Doppler broadening is described by a Voigt distribution.  When the broadening is much smaller than the transition linewidth, $\Gamma$, the Voigt distribution is well approximated to a Lorentzian with a linewidth given by (see appendix~\ref{Append2})
\begin{equation}
\Gamma'=\Gamma+\frac{4\sigma^2}{\Gamma},
\label{broadening}
\end{equation}
where $\sigma$ is the root-mean-square (rms) Doppler shift.  At the Doppler cooling limit of the transition, we estimate an increase in the measured linewidth of $\approx 11\,\mathrm{kHz}$ and this increases linearly with temperature.   For the multi-level structure involved, we can expect the temperature to be above the Doppler cooling limit and hence Doppler broadening is likely significant compared to the statistical uncertainty.  As we are currently unable to accurately determine temperature, we cannot reliably assess this systematic. However, we note that the fitted linewidths are consistent with the theoretical value of $2.61\,\mathrm{MHz}$ given in \cite{LuProperties} and any realistic temperature is unlikely to lead to a significant contradiction of that theory.
\subsection{$^3P_1$}
Frequency references for  $^1S_0\leftrightarrow{^3}P_1$ transitions are denoted $f_{F}$, where $F$ denotes the hyperfine level of $^3P_1$.  These frequencies are determined by measuring the rate at which $\ket{^1S_0,7,\pm1}$ is depopulated by a single 350-nm laser beam as a function of laser frequency.  In contrast to the $^3P_0$ measurement, we require the use of $\pi$-polarization to obtain a simple exponential decay of the $\ket{^1S_0,7,\pm1}$ population and a corresponding Lorentzian line-profile for the depumping rate.  This is due to Zeeman broadening of the upper state and an asymmetry in the excitation rates out of $\ket{^1S_0,7,\pm1}$ for the two circular polarizations which results in a sum of two displaced Lorentzians when using linear polarization orthogonal to the magnetic field direction.

The 350-nm depumping beam is also used to repump the atom into $^3D_1$ at the start of each experiment.  With pure $\pi$-polarization, $\ket{^1S_0,7,\pm7}$ and $\ket{^1S_0,7,0}$ are dark states when coupled to $F=6$ and $7$, respectively.  For the measurement of these two transition frequencies, the polarization was rotated by a few degrees to avoid long interruptions of the experiment.  For the $F=7$ case, excitation rates out of $\ket{^1S_0,7,\pm1}$ are approximately 30 times stronger for circular polarizations than for $\pi$.  This is in contrast to the $F=6$ case in which the excitation rates are smaller for circular polarizations than they are for $\pi$-polarizations.  From simulations, a $3^\circ$ rotation of the polarization results in a $\sim4\%$ increase of the linewidth for the $F=7$ case and a $<0.1\%$ increase for the $F=6$ case, but in both cases the line profile is still well described by a Lorentzian. 

The experimental sequence consists of conditional preparation to $\ket{^1S_0,7,\pm1}$, depumping with the 350-nm laser for a time $\tau$, shelving $\ket{^1S_0,7,\pm1}$ back to $\ket{^3D_1,7,0}$ with the 848-nm laser, and then detection with the 646-nm laser.  The probability that the ion is subsequently detected bright is given by
\begin{equation}
\label{3P1pumping}
p=p_\infty+(p_s-p_\infty) e^{-R \tau},
\end{equation}
where $p_s$ is the shelving efficiency and $p_\infty$ accounts for decays from $^3P_1$ to $^3D_1$ during depumping.  As in \sref{3P0}, $p$ is determined from 200 cycles of the experiment and this is repeated 50 times to determine the statistical variation, $p_s$ and $p_\infty$ are determined by independent experiments, and $R$ is found by inverting \eref{3P1pumping}.

\begin{figure}[hbt]
\begin{center}
  \includegraphics{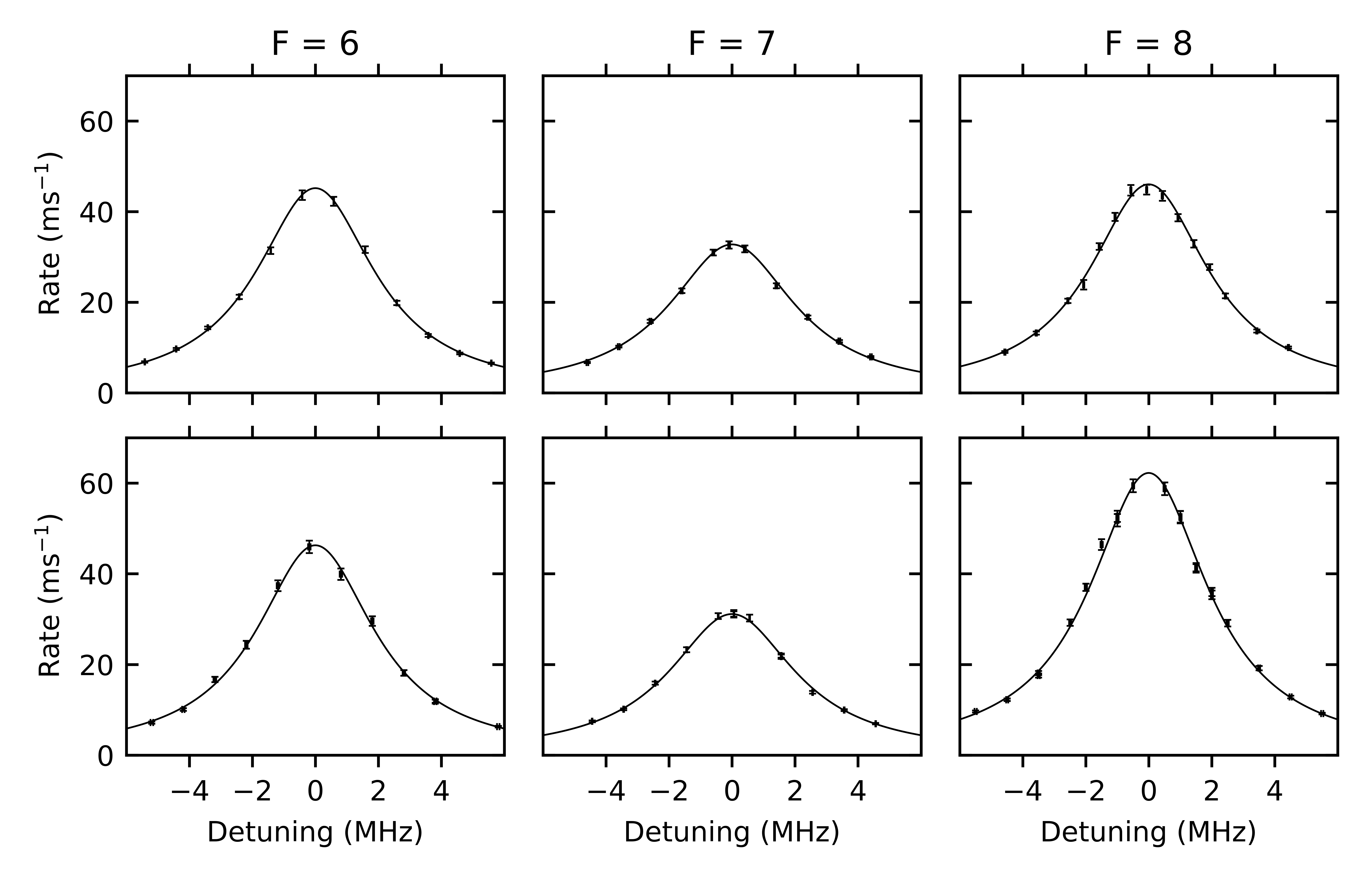}
  \caption{Line shapes for $\ket{^1S_0,7,m_F} \to \ket{^3P_1,F,m_{F}}$ with $m_F = -1$ (top row) and  $m_F = +1$ (bottom row) for $F=6$ (left), $F=7$ (middle) and $F=8$ (right). Solid curves are Lorentzian fits to the data (black dots).}
  \label{3P1Data}
\end{center}
\end{figure}

In \fref{3P1Data}, the inferred values of $R$ are plotted as a function of detuning for excitation to each of the upper $F$ levels from both $m_F=\pm 1$ ground states.  For each value of $F$, the detunings are given relative to a conveniently chosen reference point.  The solid curves are from a chi-square fit to Lorentzian profiles.  The fit constrained the linewidths of the $F=6$ and 8 profiles to be equal with an independent linewidth for the $F=7$ profiles for the reason mentioned above.  Separations of $m=\pm1$ line profiles were constrained by a single parameter proportional to $g_J \mu_B B/h$ which neglects the small dependence of the Zeeman shifts on $g_I$.   From the fit we have $g_J \mu_B B/h=4.291(88)\,\mathrm{MHz}$ which is consistent with the magnetic field determined from the ground state splitting.  The fitted linewidth for the $F=6$ and 8 profiles is $2\pi\times 4.560(33)\,\mathrm{MHz}$ compared to $2\pi\times 4.854(52)\,\mathrm{MHz}$ for $F=7$.  The difference is reasonably consistent with the expected influence of imperfect polarization.

The midpoints of the line-centers for the $m_F=\pm 1$ profiles can be taken as an estimate of the zero field line center and corresponding optical frequencies, $f_F$, determined by referencing to the comb are
\begin{subequations}
\label{fF}
\begin{align}
f_6 &= 854\, 473\, 365\, 083\, (19)\,\mathrm{kHz},\\ 
f_7 &= 854\, 500\, 434\, 764\, (18)\,\mathrm{kHz},\\
\adjintertext{0pt}{0pt}{and}
f_8 &= 854\, 526\, 227\, 250\, (15)\,\mathrm{kHz}.
\end{align}
\end{subequations}
The transfer cavity used to stabilize the 701-nm laser operates at atmosphere and hence the resonant frequency is subject to significant shifts throughout the day ($\sim 5\,\mathrm{MHz}$).  For this reason the 701-nm laser was monitored by the comb throughout the data collection, and corrections to the cavity offset made every second to maintain a given set point.  This eliminates systematic shifts associated with laser drifts and the errors given in eqs.~(\ref{fF}) are the statistical uncertainties from the fit.

As for the $^3P_0$ measurements, laser linewidth and Doppler broadening are likely to be significant compared to the statistical uncertainty in the fitted linewidths.  Comb measurements throughout the data collection have an rms deviation of $\sim30\,\mathrm{kHz}$ from the set point frequency which translates to $\sim60\,\mathrm{kHz}$ at the doubled frequency.  These deviations are normally distributed and thus, from \eref{broadening}, are unlikely to significantly influence the line profile. For the Doppler cooling limit of the $^3D_1\leftrightarrow{^3}P_0$ transition, we estimate an increase in the measured linewidth of $\approx 21\,\mathrm{kHz}$.  As for the $^3P_0$ measurements, we are unable to fully characterize this systematic.  However the fitted linewidth of  $4.560(33)\,\mathrm{MHz}$ is comparable to the theoretical value of $4.46\,\mathrm{MHz}$ given in \cite{LuProperties}.  As in the $^3P_0$ case, any realistic temperature or laser linewidth is {unlikely} to lead to a significant contradiction of the theory.

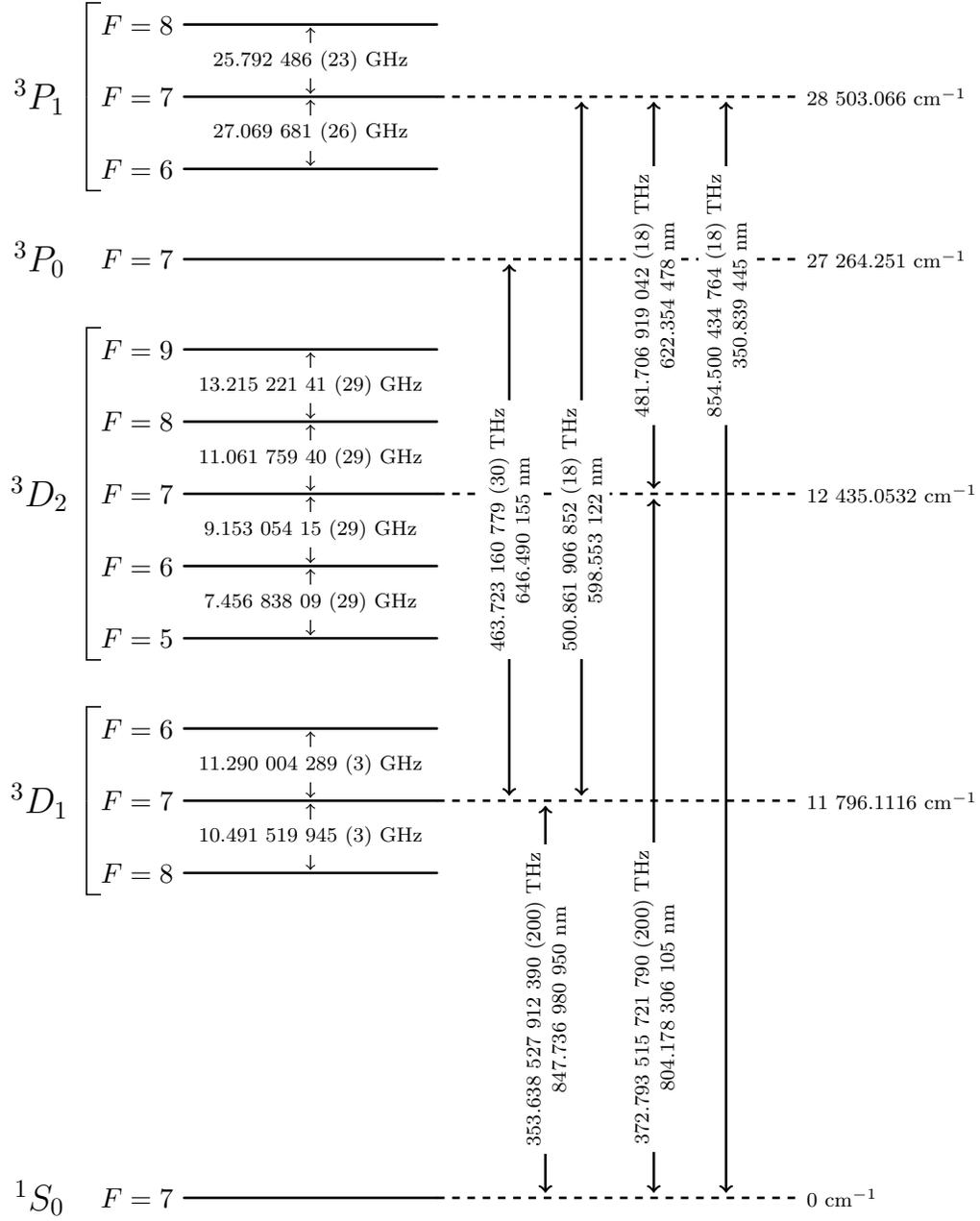
\begin{figure}[h]
\raggedright

\begin{tikzpicture}[scale=1,
      level/.style={line width=1pt,line cap=round},
      virtual/.style={thick,line width=1pt, dashed},
      guide/.style={thin,line width=0.6pt,densely dashed},
      sguide/.style={thin,line width=0.6pt},
      dimension/.style={very thick,line width=0.5pt,<->,shorten <=-3pt,shorten >=-3pt},
      transition/.style={very thick,line width=1.0pt,<->,shorten <=2pt,shorten >=2pt},
      classical/.style={line width=2pt,->,shorten <=8pt,shorten >=-3pt},
      transnode/.style={midway,rotate=90,fill=white,align=center}
    ]

    
    \coordinate (1s0) at (0cm,0cm);
    \coordinate (3d1) at (0cm,5.5cm);
    \coordinate (3d2) at (0cm,9.75cm);
    \coordinate (3p0) at (0cm,13cm);
    \coordinate (3p1) at (0cm,15.25cm);
   \def \hyperscale{0.1};
   \coordinate (midsep) at ($(5.0cm,0cm)$); 
   \coordinate (levlength) at ($(-3.5cm,0cm)$); 
   \coordinate (termlabeldim) at ($(-5.5cm,0cm)$); 
   \coordinate (transsep) at ($(1.0cm,0cm)$); 
    
\footnotesize
\draw[virtual] (1s0)--++(midsep) node[right] {0 cm$^{-1}$};
\draw[virtual] (3d1)--++(midsep) node[right] {11 796.1116 cm$^{-1}$};
\draw[virtual] (3d2)--++(midsep) node[right] {12 435.0532 cm$^{-1}$};
\draw[virtual] (3p0)--++(midsep) node[right] {27 264.251 cm$^{-1}$};
\draw[virtual] (3p1)--++(midsep) node[right] {28 503.066 cm$^{-1}$};

\normalsize

\draw[level] ($(1s0)$) --++($(levlength)$) node[left] {$F=7$} ;
\draw[level] ($(3d1)+\hyperscale*(0,-10.cm)$) -- ++ (levlength) node[left] {$F=8$} node[midway] (3d1f8) {};
\draw[level] ($(3d1)+\hyperscale*(0,0cm)$) -- ++ (levlength) node[left] (3d1b) {$F=7$} node[midway] (3d1f7) {};
\draw[level] ($(3d1)+\hyperscale*(0,10.0cm)$) -- ++ (levlength) node[left] {$F=6$} node[midway] (3d1f6) {};
\draw[level] ($(3d2)+\hyperscale*(0,-20cm)$) -- ++ (levlength) node[left] {$F=5$} node[midway] (3d2f5) {};
\draw[level] ($(3d2)+\hyperscale*(0,-10cm)$) -- ++ (levlength) node[left] {$F=6$} node[midway] (3d2f6) {};
\draw[level] ($(3d2)+\hyperscale*(0,0cm)$) -- ++ (levlength) node[left] (3d2b) {$F=7$} node[midway] (3d2f7) {};
\draw[level] ($(3d2)+\hyperscale*(0,10cm)$) -- ++ (levlength) node[left] {$F=8$} node[midway] (3d2f8) {};
\draw[level] ($(3d2)+\hyperscale*(0,20cm)$) -- ++ (levlength) node[left] {$F=9$} node[midway] (3d2f9) {};
\draw[level] ($(3p0)$) --++($1*(levlength)$)node[left] {$F=7$};
\draw[level] ($(3p1)+\hyperscale*(0,-10cm)$) --++($1*(levlength)$)node[left] {$F=6$} node[midway] (3p1f6){};
\draw[level] ($(3p1)+\hyperscale*(0,0cm)$) --++($1*(levlength)$)node[left] (3p1b) {$F=7$} node[midway] (3p1f7){};
\draw[level] ($(3p1)+\hyperscale*(0,10cm)$) --++($1*(levlength)$)node[left] {$F=8$} node[midway] (3p1f8){};


\draw[sguide] ($(3d1b)-(0.7cm,0)$) -- ++($\hyperscale*(0,13cm)$) -- ++(.2cm,0cm);
\draw[sguide] ($(3d1b)-(0.7cm,0)$) -- ++($\hyperscale*(0,-13cm)$) -- ++(.2cm,0cm);
\draw[sguide] ($(3d2b)-(0.7cm,0)$) -- ++($\hyperscale*(0,23cm)$) -- ++(.2cm,0cm);
\draw[sguide] ($(3d2b)-(0.7cm,0)$) -- ++($\hyperscale*(0,-23cm)$) -- ++(.2cm,0cm);
\draw[sguide] ($(3p1b)-(0.7cm,0)$) -- ++($\hyperscale*(0,13cm)$) -- ++(.2cm,0cm);
\draw[sguide] ($(3p1b)-(0.7cm,0)$) -- ++($\hyperscale*(0,-13cm)$) -- ++(.2cm,0cm);

\footnotesize
\draw[dimension] (3d1f8) -- (3d1f7) node[midway,fill=white] {10.491 519 945 (3) GHz}; 
\draw[dimension] (3d1f7) -- (3d1f6) node[midway,fill=white] {11.290 004 289 (3) GHz}; 
\draw[dimension] (3d2f5) -- (3d2f6) node[midway,fill=white] {7.456 838 09 (29) GHz}; 
\draw[dimension] (3d2f6) -- (3d2f7) node[midway,fill=white] {9.153 054 15 (29) GHz}; 
\draw[dimension] (3d2f7) -- (3d2f8) node[midway,fill=white] {11.061 759 40 (29) GHz}; 
\draw[dimension] (3d2f8) -- (3d2f9) node[midway,fill=white] {13.215 221 41 (29) GHz}; 
\draw[dimension] (3p1f6) -- (3p1f7) node[midway,fill=white] {27.069 681 (26) GHz}; 
\draw[dimension] (3p1f7) -- (3p1f8) node[midway,fill=white] {25.792 486 (23) GHz}; 

\footnotesize
\draw[transition] ($(1s0)+1.5*(transsep)$) -- ++ (3d1) node[transnode] {353.638 527 912 390 (200) THz\\847.736 980 950 nm};
\draw[transition] ($(1s0)+3*(transsep)$) -- ++ (3d2) node[transnode,xshift=-2.1cm] {372.793 515 721 790 (200) THz\\804.178 306 105 nm};
\draw[transition] ($(3d1)+1*(transsep)$) -- ++ ($(3p0)-(3d1)$) node[transnode] {463.723 160 779 (30) THz\\646.490 155 nm};
\draw[transition] ($(3d2)+3*(transsep)$) -- ++ ($(3p1)-(3d2)$) node[transnode] {481.706 919 042 (18) THz\\622.354 478 nm};
\draw[transition] ($(3d1)+2*(transsep)$) -- ++ ($(3p1)-(3d1)$) node[transnode,xshift=-1.125cm] {500.861 906 852 (18) THz\\598.553 122 nm};
\draw[transition] ($(1s0)+4*(transsep)$) -- ++ (3p1) node[transnode,xshift=4.875cm] {854.500 434 764 (18) THz\\350.839 445 nm};

\Large
\node[] at ($(1s0)+(termlabeldim)$) {\Large $^1S_0$};
\node[] at ($(3d1)+(termlabeldim)$) {\Large $^3D_1$};
\node[] at ($(3d2)+(termlabeldim)$) {\Large $^3D_2$};
\node[] at ($(3p0)+(termlabeldim)$) {\Large $^3P_0$};
\node[] at ($(3p1)+(termlabeldim)$) {\Large $^3P_1$};
\end{tikzpicture}
\caption{$^{176}$Lu$^+$ transition frequencies and hyperfine structures measured in this work.}
\label{frequencyChart}
\end{figure}

References for the 622-nm laser can be determined by combining measurements of $^3P_1$ together with those for $^3D_2$.  Similarly, references for a 598-nm laser, which can be used to optically pump out of $^3D_1$, can be determined by combining measurements of $^3P_1$ together with those for $^3D_1$.  A complete frequency diagram is given in \fref{frequencyChart}.

\section{Summary}
We have performed high resolution optical spectroscopy on four low-lying excited states of $^{176}$Lu$^+$. These measurements include the first observations of the $^1S_0\leftrightarrow{^3}D_1$ and  $^1S_0\leftrightarrow{^3}D_2$ clock transitions for this isotope.  Our measurements provide optical references for the 350-nm, 598-nm and 622-nm transitions, used for optically pumping out of $^1S_0$, $^3D_1$ and $^3D_2$, respectively.  Spectroscopy of the $^3D_1\leftrightarrow{^3}P_0$ transition also provides a frequency reference for cooling and detection.  The frequency references obtained are the most relevant for clock operation with $^{176}$Lu$^+$. References for the 804-nm clock transition may provide access to higher order nuclear moments depending on the accuracy at which correction terms can be calculated. All frequency references are summarized in \fref{frequencyChart}.
\section*{Acknowledgements}
We thank Rolf Persson for bringing our attention to the possible importance of additional correction terms to the HFS $D$ coefficient.
\section*{Funding}
This research is supported by the National Research Foundation, Prime Ministers Office, Singapore and the Ministry of Education, Singapore under the Research Centres of Excellence programme. It is also supported by A*STAR SERC 2015 Public Sector Research Funding (PSF) Grant (SERC Project No: 1521200080).
\bibliographystyle{tfp}
\bibliography{LuIonSpec}
\appendix
\section{Quadratic Shifts}
\subsection{$^3D_1$}
\label{3D1Zeeman}
The quadratic shifts for the $m=0$ states of $^3D_1$ are:
\begin{subequations}
\begin{align}
\Delta E_6^{(2)}/h=&+\frac{8}{15}\frac{(g_J-g_I^\prime)^2\mu_B^2 B^2}{h^2 (f_{1,6}-f_{1,7})},\\
\Delta E_7^{(2)}/h=&-\frac{8}{15}\frac{(g_J-g_I^\prime)^2\mu_B^2 B^2}{h^2 (f_{1,6}-f_{1,7})} +\frac{7}{15}\frac{(g_J-g_I^\prime)^2\mu_B^2 B^2}{h^2 (f_{1,7}-f_{1,8})},\\
\adjintertext{0pt}{0pt}{and}
\Delta E_8^{(2)}/h=& -\frac{7}{15}\frac{(g_J-g_I^\prime)^2\mu_B^2 B^2}{h^2 (f_{1,7}-f_{1,8})}.
\end{align}
\end{subequations}
In calculating the Zeeman shifts for each level, we use $g_J=0.5$ with a $0.5\%$ error to account for possible deviations from this value. We neglect any uncertainty in $g_I^\prime$, as the value itself is small compared to the uncertainty in the $g_J$, and take the value $g_I^\prime=-2.4360\times 10^{-4}$ as measured for neutral lutetium \cite{LuIgI}. Using the B field inferred from the ground state splitting, $B = 0.207 (2)\,\mathrm{mT}$, we obtain quadratic shifts:
\begin{subequations}
\begin{align}
\Delta E_6^{(2)}/h&=99.2\,(2.2)\,\mathrm{Hz},\\
\Delta E_7^{(2)}/h&=-5.80\,(0.13)\,\mathrm{Hz},\\
\adjintertext{0pt}{0pt}{and}
\Delta E_8^{(2)}/h&=-93.4\,(2.1)\,\mathrm{Hz}.
\end{align} 
\end{subequations}
\subsection{$^3D_2$}
\label{3D2Zeeman}
The quadratic shifts for the $m=0$ states of $^3D_2$ are:
\begin{subequations}
\begin{align}
\Delta E_5^{(2)}/h=&-\frac{16}{13}\frac{(g_J-g_I^\prime)^2\mu_B^2 B^2}{h^2 (f_{2,6}-f_{2,5})},\\
\Delta E_6^{(2)}/h=&+\frac{16}{13}\frac{(g_J-g_I^\prime)^2\mu_B^2 B^2}{h^2 (f_{2,6}-f_{2,5})} -\frac{102}{65}\frac{(g_J-g_I^\prime)^2\mu_B^2 B^2}{h^2 (f_{2,7}-f_{2,6})},\\
\Delta E_7^{(2)}/h=& +\frac{102}{65}\frac{(g_J-g_I^\prime)^2\mu_B^2 B^2}{h^2 (f_{2,7}-f_{2,6})} -\frac{117}{85}\frac{(g_J-g_I^\prime)^2\mu_B^2 B^2}{h^2 (f_{2,8}-f_{2,7})},\\
\Delta E_8^{(2)}/h=& +\frac{117}{85}\frac{(g_J-g_I^\prime)^2\mu_B^2 B^2}{h^2 (f_{2,8}-f_{2,7})} -\frac{14}{17}\frac{(g_J-g_I^\prime)^2\mu_B^2 B^2}{h^2 (f_{2,9}-f_{2,8})},\\
\adjintertext{0pt}{0pt}{and}
\Delta E_9^{(2)}/h=&+\frac{14}{17}\frac{(g_J-g_I^\prime)^2\mu_B^2 B^2}{h^2 (f_{2,9}-f_{2,8})}.
\end{align}
\end{subequations}
In calculating the Zeeman shifts for each level, we use $g_J=7/6$.  As with $^3D_1$ we use a $0.5\%$ error to account for possible deviations from this value and we neglect any uncertainty in $g_I^\prime$ relative to the neutral atom value. The B field inferred from the ground state Zeeman splitting,  was also found to be $B=0.207(2)\,\mathrm{mT}$.  Using this value, we obtain quadratic shifts
\begin{subequations}
\begin{align}
\Delta E_5^{(2)}/h&=-1886\,(41)\,\mathrm{Hz},\\
\Delta E_6^{(2)}/h&=-73.0\,(1.6)\,\mathrm{Hz},\\
\Delta E_7^{(2)}/h&=537.3\,(12)\,\mathrm{Hz},\\
\Delta E_8^{(2)}/h&=710.0\,(15)\,\mathrm{Hz},\\
\adjintertext{0pt}{0pt}{and}
\Delta E_9^{(2)}/h&=712.3\,(15)\,\mathrm{Hz}.
\end{align} 
\end{subequations}
Since the uncertainties are much less than the uncertainties in the optical frequency measurements, they are neglected from eqs.~(\ref{f2F}).
\section{Voigt Profile - $\sigma\ll\gamma$}
\label{Append2}
When motional frequencies are much smaller than the linewidth of the transition, the line profile in the presence of Doppler broadening is determined by the Voigt profile. The Voigt function is the convolution of a Lorentzian, with a half-width-half-maximum (HWHM) $\gamma$, and a Gaussian, with standard deviation $\sigma$.  In the limit that $\sigma \ll \gamma$, the line shape is well approximated by a Lorentzian with HWHM $\gamma'$.  To find $\gamma'$ we calculate the least squares fit of the Voigt function $V(x,\gamma,\sigma)$ to a Lorentzian by minimizing the integral
\begin{equation}
\int^\infty_{-\infty} \left(V(x,\gamma,\sigma)-\frac{\gamma' A}{\pi(\gamma'^2+x^2)}\right)^2\mathrm{d}\,x=\int^\infty_{-\infty} \left(V(\bar{x},1,\bar{\sigma})-\frac{\bar{\gamma}' A}{\pi(\bar{\gamma}'^2+\bar{x}^2)}\right)^2\mathrm{d}\,\bar{x},
\end{equation}
where we have additionally allowed for an adjustment of the amplitude and conveniently scaled by $\gamma$ with an overbar indicating the associated scaled value.  We can expand the integrand to second order in $\bar{\sigma}$ and minimize the resulting function to find $A$ and $\bar{\gamma}'$. For this, it is useful to note that
\begin{equation}
V(\bar{x},1,\bar{\sigma})\approx\frac{1}{\pi(1+\bar{x}^2)}+\frac{(3 \bar{x}^2-1)\bar{\sigma}^2}{\pi(1+\bar{x}^2)^3}.
\end{equation}
The solutions for $A$ and $\bar{\gamma}'$ can then be expanded to second order in $\bar{\sigma}$ giving
\begin{equation}
A=1+\frac{1}{4}\left(\frac{\sigma}{\gamma}\right)^2,\quad \gamma'=\gamma+\frac{\sigma^2}{\gamma}.
\end{equation}
In terms of the FWHM, we have
\begin{equation}
\Gamma'=\Gamma+\frac{4\sigma^2}{\Gamma}.
\end{equation}
\end{document}